\documentclass[prd,reprint,superscriptaddress,nofootinbib]{revtex4-1}
\usepackage{graphicx,amssymb}
\usepackage{bbm}
\usepackage{color}
\usepackage{ulem}

\newcommand{\ben}{\begin{enumerate}}
\newcommand{\een}{\end{enumerate}}
\newcommand{\bit}{\begin{itemize}}
\newcommand{\eit}{\end{itemize}}

\newcommand{\beqa}{\begin{eqnarray}}
\newcommand{\eeqa}{\end{eqnarray}}
\newcommand{\beq}{\begin{equation}}
\newcommand{\eeq}{\end{equation}}
\newcommand{\bay}{\begin{array}}
\newcommand{\eay}{\end{array}}

\newenvironment{Eqnarray}%
         {\arraycolsep 0.14em\begin{eqnarray}}{\end{eqnarray}}
\def\beqa{\begin{Eqnarray}}
\def\eeqa{\end{Eqnarray}}

\def\gsim{\ \rlap{\raise 3pt \hbox{$>$}}{\lower 3pt \hbox{$\sim$}}\ }
\def\lsim{\ \rlap{\raise 3pt \hbox{$<$}}{\lower 3pt \hbox{$\sim$}}\ }

\def\ord{{\cal O}}
\def\lt{\left}
\def\rt{\right}
\def\half{\frac{1}{2}}

\def\eg{{\it e.g.}}
\def\ie{{\it i.e.}}
\def\lag{{\cal L}}

\def\bracket#1#2 {\mathinner{\langle{#1}|{#2}\rangle}}

\begin{document}

\title{\vspace*{-0.5in} Does Supersymmetry Require Two Higgs Doublets?}

\author{Masahiro Ibe}
\affiliation{Department of Physics \& Astronomy, University of California, Irvine, CA 92697}
\affiliation{Department of Physics, University of Tokyo, Tokyo 113-0033, Japan}
\author{Arvind Rajaraman}
\affiliation{Department of Physics \& Astronomy, University of California, Irvine, CA 92697}
\author{Ze'ev Surujon}
\affiliation{Department of Physics \& Astronomy, University of California, Riverside, CA 92521}
\affiliation{Department of Physics, University of California at San Diego, La Jolla, CA 92093}

\vspace*{5mm}

\begin{abstract}
\vspace*{5mm}
We discuss a new class of low energy supersymmetric models in which
the Higgs sector includes a single doublet, for example $H_u$, but not $H_d$.
Chiral gauge anomalies are canceled against new electroweak-charged states.
We discuss the main challenges in building such models, and present several
models where these issues are addressed.
The resulting phenomenology can be distinguished from that of the MSSM
in a number of ways, most notably in physics related to down-type quarks
and charged leptons.
%
As a first step toward a chiral Higgs sector, we discuss the scenario of an inert
$H_d$ doublet.  We show that a UV completion of such model naturally includes
dark matter with novel, flavorful couplings to SM quarks.
%

\ 
\end{abstract}

\maketitle

\section{Introduction}
In the Minimal Supersymmetric Standard Model (MSSM),
the Higgs sector consists of a vector-like pair of doublets: $H_u$
and $H_d$ with hypercharges $1/2$ and $-1/2$, respectively.
Virtually all the literature of low energy supersymmetry is based on this field content
and on extensions thereof.

The need for both doublets can be justified by two arguments.
The first argument is that a single chiral doublet of Higgsinos would
suffer from gauge (and Witten) anomalies;
the second is that holomorphy prevents the up(down)-type Higgs from having
supersymmetric Yukawa couplings to down(up)-type fermions.
Both these arguments can be challenged:
anomaly cancellation can potentially be achieved with
other (perhaps chiral) matter content, and
as for holomorphy, one should realize that it is in force insofar as
supersymmetry is not broken.  Eventually, non-supersymmetric couplings
of up-type Higgs to down-type fermions will be generated at some
scale~\cite{wrong-higgs,uplifted,FMSSM}.
This suggests that it may be possible
to build supersymmetric models which do not have two vectorlike Higgs doublets.

In this note, we attempt to produce models where the Higgs sector is
chiral, \ie, the model includes the chiral superfield $H_u$, but it does not
include a field with the quantum numbers of $H_d$.
Nevertheless, these models will be anomaly-free and will have large enough masses
for down-type quarks and charged leptons.
Such models have certain very attractive features, most notably the absence
of the term $\mu H_uH_d$, thereby potentially solving the $\mu$-problem and
explaining why the Higgs is light.

The main difficulty of a chiral Higgs sector is that in the absence of $H_d$,
the Higgsino cannot acquire a mass until electroweak symmetry is broken.
Furthermore, to
cancel anomalies previously canceled by $H_d$, we will have to introduce
new fields.  The new fields must be chiral too, if they are to cancel
the anomalies, but they must be given adequately large masses in order
for the model to be phenomenologically viable.
These requirements are challenging for model-building. Nevertheless, we show
that models can be built, at least as effective field theories,
where all anomalies are canceled, and the spectrum is phenomenologically viable.

As a first step toward a chiral Higgs sector, we will present a model with
both $H_u$ and $H_d$, where $H_d$ does not
couple to quark and lepton superfields at all.  In this model, for which we also provide
an ultraviolet completion, down-type quarks and charged leptons acquire
masses without involving any component fields of $H_d$
(as opposed to~\cite{uplifted}).
Once we have established the possibility of such an inert
$H_d$, which plays no role in (s)quark and (s)lepton phenomenology, we proceed
to consider scenarios of a fully chiral Higgs sector,
where $H_d$ is entirely absent from the theory.
These models face new problems which we discuss in detail.

We analyze one such model in detail, where a single SM Higgs doublet is embedded
in a chiral fourth generation.
We show that this model is phenomenologically viable, and discuss
some consequences.
Next, we present more examples of Higgs sectors which are chiral, yet
fully massive.
We close with a discussion of future directions.

\section{Down-Type Masses Without a Down-Type Higgs}
%
In the MSSM, the Higgs doublets couple to quarks and leptons only via the
supersymmetry-preserving terms
\beq
   \int\!\! d^2\theta\lt(Y_{uij}H_uQ_i\bar u_j+Y_{dij}H_dQ_i\bar d_j
   +Y_{\ell ij}H_dL_i\bar \ell_j\rt).
   \label{eq:mssm-yuk}
\eeq
In particular, there is no tree-level coupling between $H_u$ and down-type quarks.
However, in any extension of the MSSM, additional
higher dimensional operators can be generated by new degrees of freedom at
a scale $M$.
These would generically include the terms
\beq
   \int\!\! d^4\theta \frac{X^\dagger}{M^2}\!\lt(\!
   y'_{uij}H_d^\dagger Q_i\bar u_j\!+\!
   y'_{dij}H_u^\dagger Q_i\bar d_j\!+\!
   y'_{\ell ij}H_u^\dagger L_i\bar \ell_j\!\rt),
   \label{eq:wrong-yuk}
\eeq
where $X=F_X\theta^2$ parameterizes supersymmetry breaking.
Such operators have been mentioned in~\cite{DST},
and also in~\cite{four-higgses} (where they arise by integrating out a heavy pair of
Higgs doublets).
Their presence is equivalent to having non-holomorphic terms in the superpotential,
\eg,
\beq
   \int\!\!d^2\bar\theta\,\int\!\!d^2\theta\,\frac{X^\dagger}{M^2}
   \lt(H_u^\dagger Q\bar d\rt)\equiv
   \lt(\frac{F_X}{M^2}\rt)\int\!\!d^2\theta\,\lt(H_u^\dagger Q\bar d\rt).
   \label{eq:nonholo}
\eeq
This is not surprising, since the theory is indeed explicitly supersymmetry-violating.
Effectively, the theory now includes ``wrong-Higgs'' Yukawa couplings,
\eg, between the up-type Higgs and down-type fermions~\cite{wrong-higgs}.
Such terms ostensibly induce hard
supersymmetry-breaking in the low energy Lagrangian, but being
inversely proportional to the cutoff $M$, they do not
reintroduce the hierarchy problem. 
The resulting mass matrix of the down-type quarks is
\beq
   m_d=y'_{d}v_u{F_X\over M^2}+Y_{d}v_d.
\eeq

Consider a limiting scenario where $Y_d=0$. In that case,
$H_d$ becomes ``inert'', \ie, it does not couple to SM quarks and leptons
at tree-level.
The down-type masses are generated purely by the
higher dimensional couplings to $H_u$.
The phenomenology of an inert $H_d$ doublet is different
from that of the MSSM in several ways.
For example, since the masses of the bottom and the tau are decoupled from
$v_d$, $\tan\beta$ may be very large ($v_d\ll v_u$), without being ruled out by
perturbativity (see~\cite{FMSSM} for a scenario with a similar feature).

Another distinctive feature of an inert $H_d$ scenario would be
that the Higgs decay rate to $b\bar b$ is $1/\tan^2\beta$ times its MSSM value.
Essentially, $H_u$ becomes SM-like, while $H_d$ decouples,
as far as low momentum physics is concerned.

Last, the operators in~Eq.(\ref{eq:wrong-yuk}) induce only interactions
with a Higgs scalar but never with Higgsinos.  This is a consequence
of the non-holomorphy in Eq.(\ref{eq:nonholo}).
It may have a significant effect on cascade decays, especially those which involve
third generation particles.
In the MSSM, such cascades are dominated by Higgsinos (in the form of
charginos and neutralinos), since these couple strongly to the third generation.
In the inert Higgs model, such contributions are very suppressed.
Cascade decays can therefore be used to distinguish between the MSSM and
an inert $H_d$ scenario.

\subsection*{A UV Completion}
\begin{figure}[t]
  \centering
    \includegraphics[width=0.47\linewidth]{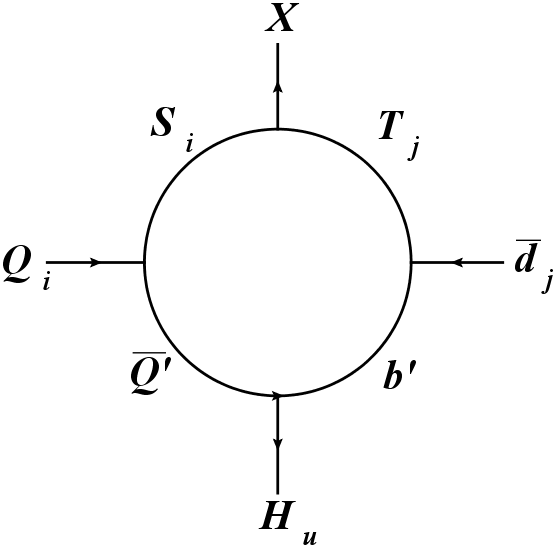}
  \caption{\footnotesize The supergraph responsible for generating
  the effective operator $X^\dagger H_u^\dagger Q_i\bar d_j$.
  \label{fig:loop}}
\end{figure}

It is possible to generate the operators in Eq.(\ref{eq:wrong-yuk})
at one loop.
As an interesting example, consider adding to the MSSM the superfields
\beq
   S_i (1,1)_0,\quad T_i(1,1)_0,\quad Q' (3,2)_{1/6}, \quad
   b' (3,1)_{-1/3},
   \label{eq:inert-content}
\eeq
along with their vectorlike counterparts
($\overline{S}_i,\overline{T}_i,\overline{Q'},\overline{b'}$),
where the index $i$ runs over three copies, and the
numbers denote the representation under the SM gauge group.
The superpotential is taken to be 
\beqa
   W&\supset& m_SS_i\overline{S}_i+m_TT_i\overline{T}_i
   +m_{Q'} Q'\overline{Q}'+m_{b'}b'\bar b'\nonumber\\
   &+&\lambda_{ij}XS_iT_j+\lambda_SS_i Q_i\overline{Q}'+
   \lambda_TT_ib'\bar d_i+H_u\overline{Q}'b'.
   \label{eq:inert-W}
\eeqa
At one loop, an effective coupling of the form
\beq
   y'_d\lt(\frac{X^\dagger}{M^2}\rt) H_u^\dagger Q\bar d
   \label{eq:effective}
\eeq
is generated (see Fig.~\ref{fig:loop}), where
\beq
   (y'_d)_{ij}\sim\lambda_{ij}\lambda_S\lambda_T/{16\pi^2},
   \label{eq:loop-operator}
\eeq
and where the scale $M$ is given by
a combination of the mass parameters in Eq.(\ref{eq:inert-W}).
The charged lepton Yukawa couplings may be generated in a similar way.
%
There are no flavor problems, since all the terms in Eq.(\ref{eq:inert-W})
are flavor universal, excepting the $\lambda_{ij}$ coupling, which generates
the entire down-type quark flavor structure. Therefore this setting is minimally
flavor violating~\cite{MFV} by construction.

It is interesting to note that in this model, the origin of the down-type flavor structure
is fully encoded in couplings between the $S_i$ and $T_i$ messengers and the
supersymmetry breaking sector.
This is in contranst to most of the literature,
where such couplings are assumed to be flavor-blind.

Another interesting feature of this model is that the superpotential~(\ref{eq:inert-W})
has a new parity.
Under the new parity, the new fields are odd, whereas the MSSM fields are even.
This renders the lightest odd particle stable, and we may choose it to be
one of the neutral $S_i$ or $T_i$, so that it would be dark matter.
Note that $S_i$ and $T_i$ carry flavor quantum numbers, potentially leading
to flavorful dark matter and interesting collider phenomenology.

In conclusion, the inert Higgs scenario can be specified both as effective field theory,
and as a full model.  Both descriptions seem to include phenomenology
which is distinct from that of the MSSM.  The particular UV completion presented
above may also include dark matter with interesting aspects for flavor physics.

\section{A Mirror Fourth Generation with a Chiral Higgs}
Having shown that the $H_d$ superfield is not necessary for
giving mass to the down-type fermions, we proceed to the more radical possibility,
namely, eliminating it from the spectrum altogether.
This is not viable by itself, since without $H_d$ the theory is anomalous.
However, including $H_d$ in the spectrum is not the only way to cancel the
anomalies due to $H_{u}$.

Let us try to replace the MSSM Higgs sector
\beq
   H_u(1,2)_\half,\quad H_d(1,2)_{-\half}
\eeq
with a new set of fields which is chiral, and thus free of the $\mu$-problem.
Perhaps the most straightforward way to do this (although not necessarily the
most minimal) is to identify $H_u$ with a lepton superfield $\overline{L'}$
of a mirror chiral fourth generation:
\beqa
   \overline{Q'}(\bar 3,2)_{-\frac{1}{6}},\, t'(3,1)_{\frac{2}{3}},\,
   b'(3,1)_{-\frac{1}{3}},\,\overline{L'}(1,2)_{\frac{1}{2}},\,
   \tau'(1,1)_{-1}. \label{eq:content}\nonumber
\eeqa
A similar idea was pursued in~\cite{GK} in the context
of low scale gravity mediation and large extra dimensions.

In order for the model to be fully chiral, we must forbid mixing between the
fourth generation, which is the new Higgs sector,
and the first three generations
\beqa
   Q_i(3,2)_{\frac{1}{6}},\, \bar u_i(\bar 3,1)_{-\frac{2}{3}},\,
   \bar d_i(\bar 3,1)_{\frac{1}{3}},\,
   L_i(1,2)_{-\frac{1}{2}},\, \bar \ell_i(1,1)_{1}.\nonumber
\eeqa
Note that such mixing is severely constrained by flavor physics in any case.
In order to do this, we impose a $Z_2$, under which only the SM generations
are odd.
In fact, this is a natural extension of the MSSM $R$-parity: the three generations
of sfermions ({\it matter} sector) are odd, whereas the new scalars, which
include also $H_u$ ({\it Higgs} sector), are even.
With these $R$-parity assignments we may still have Yukawa couplings between
the up-type Higgs and the first three generations; the up-type quarks  therefore get the usual masses.
The down type quarks and the leptons can be given masses through
non-holomorphic couplings, as described in the previous section.

We must still provide masses to the other fields of the fourth generation.
The new mirror quarks may be given mass via the terms
\beq
   y_{b'}^*\!\int\!\! d^2\theta\ \overline{Q'} b' H_u
   +   y'_{t'}\!\int\!\! d^4\theta
   \frac{X^\dagger}{M^2}H_{u}^\dagger \bar Q' t'.
\eeq
The $t'$ mass is then of order $\lt(F_X/M^2\rt)v$.
Direct collider searches at the Tevatron
already place a bound on new quark masses of $m_{q'}\gsim 350$~GeV.
Therefore we must set $F_X\sim M^2$, and the (unspecified)
UV completion is required to be such that the $y'_{t'}$ coupling is of order 1.

So far, the model conserves a ``4th generation baryon number'' $B'$,
so that either $b'$ or $t'$ is stable.
We prevent this by introducing the term
\beq
   \lambda\!\int\!\! d^2\theta\ \lt(b'\bar\tau\bar t\rt),
\eeq
such that the new quarks decay via $\tau \tilde t$ or $\tilde\tau t$.
This term introduces (non-minimal) flavor violation, but it does not conflict with
existing experimental data, since it only involves the relatively poorly measured
top and tau.

Another term allowed by $R$-parity is $Q_iQ_jb'$, but we must forbid it so that
the $b'$ would not mediate fast proton decay.
This can be done using discrete symmetries,
such as ``baryon parity'', under which all the quark fields (including the fourth
generation) are odd while other fields are even.%

So far, the Higgsinos and the $\tau'$ fermion are yet to be given mass.
A superpotential term of the form $H_uH_u\tau'$ is identically vanishing,
due to SU$(2)_L$
gauge invariance, but if we introduce an SU(2) triplet $\phi(1,3)_0$,
we can match its fermionic degrees of freedom with those of $H_u$ and $\tau'$,
via the terms
\beq
   \lambda_H\int\!\!d^4\theta
   \frac{X^\dagger H_u^\dagger \phi H_u}{M^2}
   +\lambda_{\tau'}\int\!\!d^2\theta\frac{H_u^T\phi H_u\tau'}{m}+{\rm c.c.}\ .
   \label{eq:chargino}
\eeq
We take $\phi$ to be neutral under the $R$-symmetry in order to
forbids a mass term of the form $\phi^2$
(a small mass will be generated at higher orders since $R$-symmetry is not exact, but
will be suppressed by loop factors).
This results in the mass terms
\beqa
   \lag&\supset&\lt(\lambda_H v\frac{F}{M^2}\rt)\widetilde{H}_u^+\widetilde{\phi}^-
   -\lt(\frac{\lambda_{\tau'}v^2}{m}\rt)\widetilde{\phi}^+\tau'^-\nonumber\\
   &&-\lt(\frac{\lambda_H v}{\sqrt{2}}\frac{F}{M^2}\rt)
   \widetilde{H}_u^0\widetilde{\phi}^0+{\rm c.c.}\ .
\eeqa
All fermions thus become massive.
Since one of the mass terms above is of order $v^2/m$, the parameter $m$
is required to be at the order of the weak scale in this case.%
\footnote{%
Here, we have assumed that the newly added $\phi$ does not develop 
a vacuum expectation value.
This assumption is compatible with radiative electroweak symmetry
breaking, since $H_u$ obtains a negative  mass squared
due to radiative corrections from the large Yukawa couplings
discussed above, but $\phi$ does not.}

So far, we have discussed masses for the  fermions. The new scalars get soft
supersymmetry breaking masses normally, through couplings of the form
\beq
   \kappa{X^\dagger X\over M^2}\Phi^\dagger\Phi,
\eeq
where $\Phi$ denotes collectively standard model chiral superfields.
This operator induces scalar masses at the order of
$m_0\sim \sqrt{\kappa}(F_X/M)$,
which is parametrically larger than the fermion masses by a factor of $(M/v)$,
thereby reintroducing the hierarchy problem.
However, this problem need not occur if $X$ is part of a strongly coupled sector
as in models of conformal sequestering~\cite{conformal-seq}.
In such theories , it is possible for $X^\dagger X$ to have a large anomalous
dimension in such a way that the dimension of $X^\dagger X$ is $d\simeq  1$.
The  operator is then suppressed by an additional factor $(\Lambda/v)^{d-2}$, where
$\Lambda$ is the scale of strong coupling in this sector.
The scalar masses would then be of order
$m_0\sim\lt({F_X\over M}\rt)\lt({\Lambda\over v}\rt)^{\frac{d-2}{2}}$,
which can make the scalar and fermion masses again of the same order.%
\footnote{
Instead of invoking the chiral sequestering mechanism,
we may admit a hierarchical mass spectrum between 
the gaugino and sfermions \`{a} la split supersymmetry~\cite{ArkaniHamed:2004fb},
which is naturally realized in supergravity mediation  without singlet
supersymmetry breaking fields~\cite{Giudice:1998xp}.
There, the sfermion masses are generated at tree level
and are of the order of the gravitino mass, $m_{3/2}=\ord(100)$\,TeV,
while the gaugino masses are radiatively generated by anomaly
mediation effects.
In these cases, we may trade the supersymmetry breaking spurion $X$ for
the chiral compensator $\phi = \lt(1+m_{3/2}\theta^2\rt)$,
along with the condition $M\simeq m_{3/2} = \ord(100)$\,TeV. 
}
All new fields therefore acquire a mass of order the electroweak scale, rendering
the model phenomenologically viable. As in the MSSM, the Higgs mass term gets
large negative contributions to its RGE running from top loops,
but now it will get even larger contributions from
the new heavy fermions.
This naturally drives electroweak symmetry breaking in these theories.

The phenomenology of this model is very different from the MSSM.
It is in many ways similar to a supersymmetric theory with four generations;
experimentally, the quark sector would behave exactly like the usual fourth
generation theory.
The ``lepton'' sector is more interesting, since the lepton doublet is missing; it has
become the Higgs.
However,  
we have introduced a new field
$\phi$ which 
includes one neutral and two charged degrees of freedom.
 The spectrum therefore includes one extra ``chargino'' compared to the MSSM.
Furthermore, some fields in the leptonic sector are
expected to be quite light, which may lead to interesting signals~\cite{Carpenter:2010sm}.

Also, the couplings of the fields in the ``lepton'' sector are in general quite different, and can
potentially be used to distinguish these scenarios. For example,
the neutral scalar fields in this sector are the Higgs scalar and $\phi^0$. This
is to be compared to the usual fourth generation scenario which has the
sneutrino as the neutral field.
However, the Higgs scalar couples strongly to the quarks, in particular the top quark and the new fourth
generation quarks, while the sneutrino does not have a strong interaction with quarks.
This would allow us to distinguish the chiral Higgs scenario from the usual fourth generation scenario, for
example by 
{measuring} the couplings between the fourth generation quarks and leptons. The study of these
possibilities would
be model dependent, since it is sensitive to the precise spectrum of the
theory. We shall leave this for future work. 

\section{Other  Models with a Chiral Higgs Sector}
\label{sec:colorless}
A fourth generation is not the only possibility, and
there are other candidates for a chiral Higgs sector.
It would be particularly interesting to find models which include no extra colored
particles, thereby avoiding the stringent bounds from Tevatron and LHC.
Two main challenges which are common to all chiral Higgs models are:
avoiding fractionally charged particles (these would be stable and thus severely
constrained by cosmology), and giving large enough mass to all new fermions,
especially the charged ones.
Our requirements from any chiral Higgs sector are that:
\bit
   \item It includes $H_u(1,2)_{1/2}$, and does not include $H_d(1,2)_{-1/2}$
   \item It is anomaly free
   \item It does not include any light charged particles
   (the bounds are roughly around $\sim 100$~GeV, from LEP)
   \item It does not include new stable charged states, implying
   no fractionally charged fields in the model.
   This requirement may be relaxed in scenarios with low scale reheat temperature.
\eit

The following are simple examples of models which satisfy all these requirements,
including the absence of fractionally charged fields:
\ben
   \item
      $T(1,3)_{-1},\ H_u(1,2)_{1/2},\ D(1,2)_{7/2},\
      \rho(1,1)_{2},\\ \sigma(1,1)_{-3},\  \omega(1,1)_{-4}$:\\ \ \\
      The Lagrangian is given by
      \beqa
         \delta\lag &=&\int\!\!\! d^2\theta \lt(\lambda H_uH_u T+y H_uD\omega\rt)\nonumber\\
         &&+\int\!\!\! d^4\theta\,\frac{X^\dagger}{M}\!\!\lt(\!\frac{\lambda' H_u^\dagger
         H_u^\dagger T \rho}{m}+y' H_u^\dagger D\sigma\!\rt)\!.
      \eeqa
      This gives rise to the mass terms
      \beqa
         \delta\lag_M&=&\lambda v\lt(T^0\widetilde{H}_u^0-\sqrt{2}\widetilde{H}_u^+\rt)
         +y v D^{+4}\omega^{-4}\nonumber\\
         &&+\frac{\lambda'v^2}{m}\frac{F_X}{M^2}T^{-2}\rho^{+2}
         +y'v\frac{F_X}{M^2}D^{+3}\sigma^{-3}.
      \eeqa
   \item
      $3\times\mbox{\Large \{}H_u^i(1,2)_{1/2},\ \tau'^i(1,1)_{-1},\ 
      N^i(1,1)_0\mbox{\Large \}},\ \phi(1,3)_0,\\ D(1,2)_{-3/2},\ \bar\tau'(1,1)_1,\ \rho(1,1)_2$:\\ \ \\
      {\raggedright This is a {\it chiral three-Higgs doublet model}.
      The Lagrangian is given by}
      \beqa
        \delta\lag &=&\int\!\!d^2\theta\lt(\frac{\lambda_{ijk}}{m}H_u^iH_u^j\phi\tau'^k%
         +y_{ijk}H_u^iH_u^j\tau'^k\rt.\nonumber\\
         &&+z_iH_u^i D\bar\tau'\mbox{\huge )}
         +\int\!\!d^4\theta\frac{X^\dagger}{M}\mbox{\huge (}
         \lambda'_{ij}H_u^{\dagger i}
         \phi H^j\nonumber\\
         &&+y'_{ijk}H_u^{\dagger i}H_u^jN^k
         +z'_iH_u^{\dagger i}D\rho\mbox{\huge )}.
      \eeqa
   \item
      $H_u(1,2)_{1/2},\ \phi(1,3)_0,\ T(1,3)_{-1},\ \Psi(1,4)_{1/2},\\
             D(1,2)_{-3/2},\ \bar\tau'(1,1)_1,\ \rho(1,1)_2$:\\ \ \\
      The Lagrangian is
      \beqa
         \delta\lag&=&\int\!\!d^2\theta\lt(\lambda_1H_uH_uT+\lambda_2
      H_uT\Psi+yH_uD\bar\tau'\rt)\nonumber\\
      &&\!\!\!+\!\!\int\!\! d^4\theta \frac{X^\dagger}{M}\!\lt(
            \lambda'_1 H_u^\dagger\phi H_u\!+\!\lambda'_2 H_u^\dagger\phi\Psi
            \!+\!y' H_u^\dagger D\rho\rt)\!.
      \eeqa
\een
The field contents above are selected examples and by no means constitute
an exhaustive list.
One general feature, however, seems to be the existence of
doubly, and in some cases also triply and quadruply charged particles.
Models based on such spectra may lead to novel decays in colliders.

\section{Discussion}
We have discussed models which are supersymmetric but include
only one Higgs doublet.
As a first step, we presented a model in which the $H_d$ superfield is inert
and does not contribute to SM fermion masses.
This model was analyzed both as an effective field theory, and as a UV complete
theory which may include flavorful dark matter with novel phenomenology.

Next, we discussed models with {\it chiral Higgs sectors}.
These models include only one Higgs doublet, but are nevertheless
anomaly free and phenomenologically viable. All the
new particles acquire masses at the order of the electroweak scale. 
%

Consider electroweak corrections that affect the $S$ and $T$ parameters. Since we
have a large number of new particles, these corrections may
be large. It would be interesting to see if any of the models above can be compatible
with precision electroweak tests.

Furthermore, these models in general should have striking signatures at the LHC.
This is because anomaly cancellation requires the existence of several new chiral
superfields,  which will necessarily have masses
at the electroweak scale.  Moreover, we find that the new fields are either colored, or
have unusual charges, leading to a rich phenomenology at the LHC.
Especially, the production cross section of the Higgs boson
can be highly enhanced in those models with a mirror fourth
generation.
For example, a Higgs boson mass in the range of
$120~{\rm GeV} < m_h < 600~{\rm GeV}$
has been excluded by CMS at 95\% C.L.~\cite{CMS-internal} in
models with the fourth generation.
Therefore, the idea of a mirror family can be partially tested in near future 
via the Higgs search (see for example~\cite{randy,linda} for a recent study). 

In all, the detailed investigation of chiral Higgs phenomenology appears to be
model dependent, since, as we have shown, there are many possibilities for such
models - with qualitatively different field contents.
It may however be possible to find generic features of these models, which
may allow us to distinguish the chiral Higgs theories from the usual extensions
of the MSSM. We hope to return to these questions in future work.

\acknowledgments
We would like to thank Bogdan Dobrescu, Jonathan Feng and Yuri Shirman
for valuable discussions.
Z.S. thanks Hitoshi Murayama for his insights and moral support at early stages of this work.
The work of A.R. is
supported in part by NSF Grants No. PHY-0653656 and PHY-0709742.
Z.S. was supported in part by DOE grants FG03-97ER40546 (UCSD)
and FG03-94ER40837 (UCR).

\end{document}